\documentclass[aps,pre,twocolumn,groupedaddress,showkeys,longbibliography]{revtex4-2}
\usepackage{amsmath,amsthm,amssymb}
\usepackage{braket}
\usepackage[pdftex]{graphicx}
\usepackage{mathtools}

\newcommand{\tr}{\operatorname{Tr}}

\renewcommand{\H}{\mathcal{H}}

\begin{document}

\title{Switching the function of the quantum Otto cycle in non-Markovian dynamics: heat engine, heater and heat pump}

\author{Miku Ishizaki*}
\affiliation{Department of Physics, The University of Tokyo, 5-1-5 Kashiwanoha, Kashiwa-shi, Chiba 277-8574, Japan}
\email[]{miku-ish@iis.u-tokyo.ac.jp}

\author{Naomichi Hatano$^\dagger$}
\affiliation{Institute of Industrial Science, The University of Tokyo, 5-1-5 Kashiwanoha, Kashiwa-shi, Chiba 277-8574, Japan}
\email[]{hatano@iis.u-tokyo.ac.jp}

\author{Hiroyasu Tajima$^\ddagger$}
\affiliation{Graduate School of Informatics and Engineering, The University of Electro-Communications, 1-5-1 Chofugaoka, Chofu-shi, Tokyo 182-8585, Japan}
\affiliation{JST PRESTO, 4-1-8 Honcho, Kawaguchi-shi, Saitama 332-0012, Japan}
\email[]{hiroyasu.tajima@uec.ac.jp}

\begin{abstract}
Quantum thermodynamics explores novel thermodynamic phenomena that emerge when interactions between macroscopic systems and microscopic quantum ones go into action.
Among various issues, quantum heat engines, in particular, have attracted much attention as a critical step in theoretical formulation of quantum thermodynamics and investigation of efficient use of heat by means of quantum resources.
In the present paper, we focus on heat absorption and emission processes as well as work extraction processes of a quantum Otto cycle.
We describe the former as non-Markovian dynamics, and thereby find that the interaction energy between a macroscopic heat bath and a microscopic qubit is not negligible.
Specifically, we reveal that the interaction energy is divided into the system and the bath in a region of the short interaction time and remains negative in the region of the long interaction time.
In addition, a counterintuitive energy flow from the system and the interaction energy to the hot bath occurs in another region of the short interaction time.
We quantify these effects by defining an index of non-Markovianity in terms of the interaction energy.
With this behavior of the interaction energy, we show that a non-Markovian quantum Otto cycle can switch functions such as an engine as well as a heater or a heat pump by controlling the interaction time with the heat bath.
In particular, the qubit itself loses its energy if we shorten the interaction time, and in this sense, the qubit is cooled through the cycle.
This property has a possibility of being utilized for cooling the qubits in quantum computing.
We also describe the work extraction from the microscopic system to a macroscopic system like us humans as an indirect measurement process by introducing a work storage as a new reservoir.
\end{abstract}

\keywords{Quantum thermodynamics, Quantum measurement, Non-Markovian dynamics}

\maketitle

\section{Introduction}
\quad
Quantum thermodynamics, which extends the framework of thermodynamics to microscopic scales, is attracting much attention.
It explores novel macroscopic phenomena that microscopic quantum effects cause in interactions between macroscopic systems, such as heat baths and reservoirs, and microscopic systems, such as qubits and quantum oscillators \cite{Vinjanampathy16, Deffner2020, Potts2019}.
In formulating the theory of quantum thermodynamics, as in the case of conventional thermodynamics, the study of heat engines is fundamental \cite{Mukherjee2021}. 
Here, we study a quantum heat engine, aiming to construct the theory of quantum thermodynamics.

\quad
The quantum heat engine was first noted in the field of quantum optics; mechanism of laser emission was quantum mechanically described as a heat engine \cite{Scovil59,Geusic67}. 
In the quantum heat engine, the macroscopic working substance, \textit{e.g.}\ gas inside a piston, in conventional heat engines is replaced with a quantum system. 
For the purpose of revealing thermodynamic features, the method of open quantum systems are often used \cite{Alicki1979, Geva96, Kosloff13, Quan07, Quan09, Udo16, Geva1992, Kieu2006, Li2017}.
As an example of macroscopic manifestation of quantum nature, the effect of quantum coherence \cite{Scully2003, Quan2006, Gelbwaser-Klimovsky2015, Niedenzu2015, Turkpence16,Tajima2021} and the effect of quantum entanglement \cite{Dillenschneider2009,  Zhang2008, Zhang2007, Wang2009, Thomas2011, Altintas14, kamimura2021quantum, Campisi2016, Hardal2015} were explored.
As a general theory of quantum heat engines, a trade-off relationship between the time required for one heat engine cycle and efficiency has been also derived \cite{PhysRevE.96.022138}.
Experiments on the quantum effects have been conducted \cite{Dawkins2016,Ono2020,PhysRevLett.109.203006,PhysRevLett.112.030602, PhysRevB.94.184503, Campisi2015, Josefsson2018, Fialko2012, Bergenfeldt2014, Zhang2014}.

\quad
Research on quantum heat engines has attracted much attention not only for formulating the theory of quantum thermodynamics as we describe above, but also for investigating efficient use of heat by quantum resources.
From the viewpoint of quantum information theory, work extraction by quantum measurement \cite{Elouard17} and a new heat engine based on quantum measurement \cite{Buffoni2019} were proposed.

\quad
In the study of quantum heat engines, there are several issues to be solved: how to describe interactions between microscopic and macroscopic systems; how to employ quantum correlations; how to utilize feedback of information from quantum measurements; and so on.
In the first issue, in particular, we are interested in the manifestation of microscopic properties in macroscopic systems.
In the present study, we focus on two aspects of this issue: heat exchange and work extraction.

\quad
Let us first discuss the heat exchange between microscopic and macroscopic systems.
In the conventional framework of thermodynamics, the interactions exist only among macroscopic systems, such as gas inside a piston and heat baths.
Since the interaction takes place at the surface between macroscopic systems, the dimensionality of the interaction is reduced by one, and hence the interaction can be safely ignored in conventional thermodynamics.
However, interactions among microscopic and macroscopic systems may not be treated in this way.

\quad
The interaction dynamics is quite often described by the Markov approximation  \cite{Gorini76, Lindblad1976, Breuer2002, Rivas2012}, which is valid when the time scale of the heat bath is much shorter than the time scale of the evolution of the quantum working substance.
It was revealed \cite{Shirai21} that when the heat exchange of a quantum Otto cycle is described as a Markov process, the expectation value of the interaction energy vanishes, and hence the standard thermodynamics straightforwardly extends into the quantum region.

\quad
Except for such a special case, however, we should treat the dynamics as a non-Markovian process \cite{Breuer09, Rivas10, Chru14, Bylicka14, Hashimoto19}.
For instance, several quantum heat engines are discussed from the viewpoint of non-Markovian dynamics \cite{Gelbwaser13, Zhang_2014, Kutvonen15, Bogna16, Pezzutto_2016, Raam16, Strasberg_2016, Samyadeb17, Marcantoni17, Chen17, Hamedani18, Thomas18, Wiedmann_2020,Carrega2022,Cangemi2020}.
The expectation value of the interaction takes a finite value when the process is described as a non-Markov one \cite{Shirai21}, and hence we need nontrivial generalization of thermodynamics in the quantum regime when the non-Markovianity is strong in the process in question.

\quad
In the present study, we calculate the heat exchange in the quantum Otto cycle in terms of a non-Markovian time-convolutionless master equation \cite{Guarnieri16, Hashitsume1977}.
We reveal that the interaction energy is divided into the system and the bath in a region of the short interaction time and remains negative in the region of the long interaction time.
In addition, a counterintuitive energy flow from the system and the interaction energy to the hot bath occurs in another region of the short interaction time.
We quantify these effects by defining an index of non-Markovianity in terms of the interaction energy.
Moreover, we show that counterintuitive behavior, such as a heater and a heat pump which convert work to a heat flow, appears instead of a heat engine in some parameter regions.
In particular, the qubit itself loses its energy if we shorten the interaction time between the qubit and the bath. In this sense, the qubit is cooled through the cycle, which property could be used for cooling the qubits in quantum computing.

\quad
We next discuss the work exchange between a microscopic system and a macroscopic system.
The work done by a microscopic system has been conventionally defined as the change in the expectation value of the Hamiltonian of the system of interest \cite{Vinjanampathy16, Deffner2020, Potts2019,Tasaki00,Anders13}.
However, if the work done by a quantum system is to be eventually harvested by a macroscopic system, such as us humans, it is necessary to model the work extraction using quantum measurements \cite{Hayashi17}.
In the present study, we develop a model of work extraction in a quantum Otto cycle in terms of a quantum measurement of a work storage.

\quad
The paper is organized as follows.
In Sec.~II, we present the model of a quantum Otto cycle consisting of a qubit as the working substance, two Boson heat baths, and two work storages.
We analyze the cycle in terms of non-Markovian heat absorption and emission processes as well as work extraction processes as quantum measurements, the latter of which we specifically describe in Sec.~III.
In Sec.~IV, we discover that the non-Markovian quantum Otto cycle behaves not only as an engine but also as a heat pump and a heater depending on the interaction duration between the qubit and each heat bath.
We define an index of the non-Markovianity in terms of the energy changes.
We summarize our conclusions and perspectives in Sec.~V.

\section{Model}
\label{Model}
\begin{figure}[h]
\includegraphics[width=0.5\textwidth]{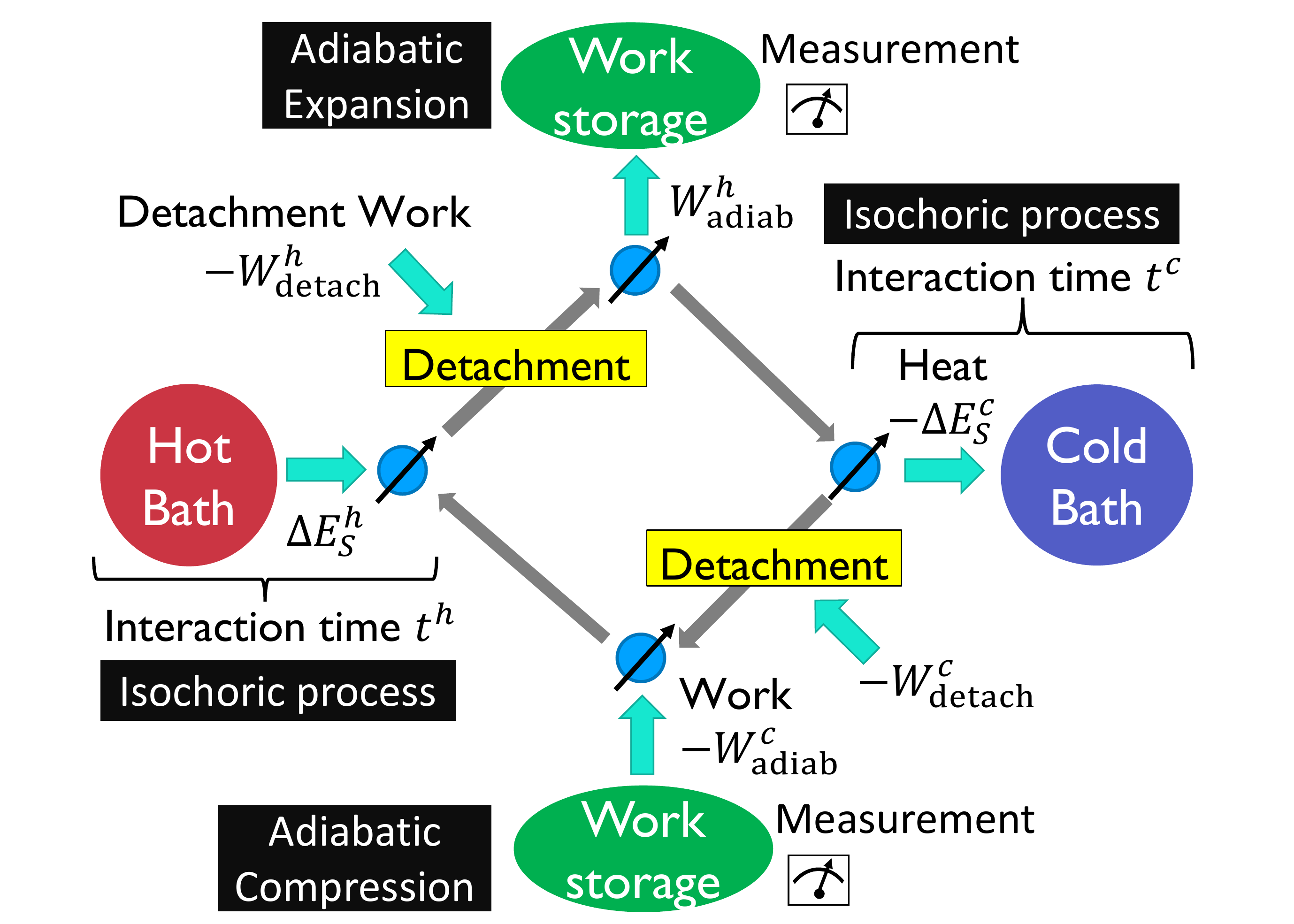}
\caption{A schematic illustration of our quantum Otto cycle. A qubit absorbs or emits heat from or to the bath in each isochoric process and does work to and receive work from the work storage in the adiabatic processes. After each isochoric process, we need detachment work to separate a qubit from the bath. We also perform measurement in order to quantify the extracted work in each adiabatic process. We define a positive energy exchange always as an energy flow into the qubit and a positive work as an energy flow out of the qubit.}
\label{otto_cycle}
\end{figure}

\quad
In the present paper, we use the quantum Otto engine analyzed in Ref.~\cite{Shirai21} for the heat exchange processes and update the work extraction processes following Ref.~\cite{Hayashi17}. 
The former are isochoric processes and the latter are adiabatic processes of one qubit as the working substance (Fig.~\ref{otto_cycle}).
In the isochoric processes, the qubit exchanges heat $\Delta E_s^h$ and $\Delta E_s^c$ with the hot and cold baths, respectively; we define a positive energy exchange always as an energy flow into the qubit.
The qubit also gives and takes the work to and from the work storages in adiabatic compression and expansion processes, respectively; we define a positive work as an energy flow out of the qubit.
After each isochoric process, we perform a detachment work, which we will explain later.
We repeat this cycle until it reaches a limit cycle.

\quad
For the isochoric processes, we use the spin-Boson model given by the Hamiltonian
\begin{align}
\H^\mu&=\H_S^\mu+\H_B^\mu+\H_I^\mu,
\end{align}
where $\mu=h$ or $c$ for the interaction with hot or cold bath, respectively, with
\begin{align}
&\H_S^\mu=\frac{\omega_0^\mu}{2}(\sigma_z+1)\:,\\
&\H_B^\mu=\sum_k\epsilon_k^\mu {b^\mu_k}^\dagger b^\mu_k\:,\\
&\H_I^\mu=\sigma_x\otimes\sum_k\left(g_k^\mu b_k^\dagger+g_k^{\mu *} b_k\right)\:,
\end{align}
where we set $\hbar=1$. Here, $\sigma_z$ and $\sigma_x$ are the Pauli operators which represent the qubit and $\omega_0^\mu\;\;(\mu=h\;\mathrm{or}\;c)$ is its Larmor frequency, which is changed during the adiabatic expansion from $\omega_0^h$ to $\omega_0^c$ and during the adiabatic compression from $\omega_0^c$ to $\omega_0^h$ with $\omega_0^h>\omega_0^c$. On the other hand, $b^\mu_k$ and ${b^\mu_k}^\dagger$ are the Boson creation and annihiration operators of mode $k$ with energy $\epsilon^\mu_k$ of hot ($\mu=h$) or cold ($\mu=c$) bath.
In the interaction Hamiltonian $H_I^\mu$, the parameter $g_k^\mu$ denotes the interaction strength between the qubit and the hot ($\mu=h$) or cold ($\mu=c$) bath of mode $k$.
We set the Ohmic spectral density function
\begin{align}
J^\mu(\omega)&=\sum_k |g_k^\mu |^2\delta(\omega-\epsilon^\mu_k)\\
&=\lambda^\mu\omega\exp(-\omega/\Omega^\mu)
\end{align}
for the qubit-bath interaction, where $\lambda^\mu$ is the coupling constant and $\Omega^\mu$ is the cutoff frequency.
For brevity, we will drop the superscript $\mu$ hereafter when common equations apply to both baths.

\quad
For an accurate description of the interaction dynamics between the qubit and the bath in each isochoric process, we employ the time-convolution-less (TCL) master equation;
\begin{align}
\frac{d}{dt}\rho_{m}^\mu(t)&=\Xi(t)\rho_{m}^\mu(t),
\end{align}
where $\rho_{m}^\mu(t)\equiv\mathrm{Tr_B}\left[\rho_{\mathrm{tot}}^\mu(t)\right]$ indicates the partial trace over the bath on the state of the total system and $\Xi(t)$ is a superoperator.
We assume the initial state for the interaction dynamics with a bath to be a product state $\rho_{\mathrm{tot}}^\mu=\rho_m^\mu\otimes\rho_B^{\mu,\mathrm{eq}}$, where the qubit state is $\rho_m^\mu\equiv\ket{m}\bra{m}$ $(m=0$ or $1)$, while the bath is in the thermal equilibrium state $\rho_B^{\mu,\mathrm{eq}}\equiv\mathrm{exp}[-\beta^\mu \H_B^\mu]/\mathrm{Tr}[\mathrm{exp}[-\beta^\mu \H_B^\mu]]$ at the inverse temperature $\beta^\mu$.
In the present paper, we employ the TCL master equation up to the second order of the interaction Hamiltonian $\H_I^\mu$ \cite{Guarnieri16}, in which the qubit follows the non-Markovian dynamics given by \cite{Guarnieri16, Shirai21}:
\begin{align}
\label{master_eq}
\frac{d}{dt}\rho_{m}^\mu(t)&=-i\left[\H_S^\mu,\rho_{m}^\mu(t)\right] \notag \\
&-\int^t_0 d\tau \mathrm{Tr_B}\left[\H_I^\mu,\left[\H_I^\mu(-\tau),\rho_{m}^\mu(\tau)\otimes\rho_B^{\mu,\mathrm{eq}}\right]\right],
\end{align}
where the interaction Hamiltonian is put in the interaction picture: $\H_I^\mu(t)=\mathrm{exp}[i\H_0^\mu t]\;\H_I^\mu\;\mathrm{exp}[-i\H_0^\mu t]$ with $\H_0^\mu\equiv\H_S^\mu+\H_B^\mu$.

\quad
After each isochoric process based on the TCL master equation for time $t^\mu$ under the interaction with the $\mu$th bath from the initial state $\ket{m}\bra{m}\otimes\rho_B^{\mu,\mathrm{eq}}$ ($m=0,1$), the ($\nu,\nu$) element of the qubit's state $\rho_{m,\nu\nu}^\mu(t^\mu)$ become \cite{Guarnieri16}
\begin{align}
\rho_{m,00}^\mu(t^\mu)=&e^{\int_0^{t^\mu}d\tau a(\tau)}\times \notag \\
&\left(\rho_{m,00}^\mu(0)-\int_0^{t^\mu}d\tau \;b(\tau)e^{-\int_0^\tau dsa(s)}\right)\label{qubitstate},\\
\rho_{m,11}^\mu(t^\mu)=&1-\rho_{m,00}^\mu(t^\mu)
\end{align}
with
\begin{align}
a(\tau)=&-2\int_0^\tau du\;D_1(u)\mathrm{cos}(\omega_0u),\\
b(\tau)=&-\int_0^\tau du \left[\Phi(u)e^{i\omega_0 u}+\Phi(-u)e^{-i\omega_0 u}\right],\\
\Phi(\tau)=&\frac{1}{2}\left(D_1(\tau)-iD_2(\tau)\right),\\
D_1(\tau)=&2\int_0^\infty d\omega J(\omega)\mathrm{coth}\left(\frac{\omega}{2 T_B}\right)\mathrm{cos}(\omega \tau)\\
=&2\lambda\Biggl(\Omega^2\frac{(\Omega\tau)^2-1}{((\Omega\tau)^2+1)^2}\notag\\
&+2T_B^2\:\mathrm{Re}\left[\psi'\left(\frac{T_B(1+i\Omega\tau)}{\Omega}\right)\right]\Biggr),\\
D_2(\tau)=&2\int_0^\infty d\omega J(\omega)\mathrm{sin}(\omega \tau)\\
=&\frac{4\lambda\Omega^3\tau}{(1+\Omega^2\tau^2)^2}.
\end{align}
Here, $D_1(\tau)$ is a noise kernel, $D_2(\tau)$ is a dissipation kernel, and $\psi'$ is the Euler digamma function $\psi'(z)\equiv\Gamma'(z)/\Gamma(z)$.
We note that we have numerically confirmed the positivity of the qubit's density matrix under the non-Markovian dynamics for all the results in the present paper.

\quad
We analyze this cycle after it converges to a limit cycle \cite{Shirai21}.
Let $P_n^\mu$ denote the probability that the qubit's state is $\ket{0}\bra{0}$ before it interacts with the $\mathrm{hot}\;(\mu=h)\;\mathrm{or}\;\mathrm{cold}\;(\mu=c)$ bath at the time of the $n$th interaction.
The probabilities satisfy the following relationships:
\begin{align}
\label{probability_c_n}
P^c_n&=P^h_n\rho^h_{0,00}(t^h)+(1-P^h_n)\rho^h_{1,00}(t^h)
\end{align}
and
\begin{align}
\label{probability_h_n}
P^h_{n+1}&=P^c_n\rho^c_{0,00}(t^c)+(1-P^c_n)\rho^c_{1,00}(t^c).
\end{align}
By taking the limit $n\rightarrow\infty$ in eqs.~(\ref{probability_c_n}) and (\ref{probability_h_n}) to obtain a limit cycle, we find \cite{Shirai21}
\begin{align}
\label{large_p}
P^\mu(t^h,t^c)&\equiv\lim_{n \to \infty}P^\mu_n=\frac{p^\mu(t^h,t^c)}{1-p_0(t^h,t^c)},
\end{align}
where
\begin{align}
\label{mini_p}
p_0(t^h,t^c)&\equiv\left[\rho_{0,00}^c(t^c)-\rho_{1,00}^c(t^c)\right]\left[\rho_{0,00}^h(t^h)-\rho_{1,00}^h(t^h)\right],\\
\label{mini_ph}
p^{h}(t^h,t^c)&\equiv\rho_{0,00}^{c}(t^{c})\rho_{1,00}^{h}(t^{h})+\rho_{1,00}^{c}(t^{c})\rho_{1,11}^{h}(t^{h}),\\
\label{mini_pc}
p^{c}(t^h,t^c)&\equiv\rho_{0,00}^{h}(t^{h})\rho_{1,00}^{c}(t^{c})+\rho_{1,00}^{h}(t^{h})\rho_{1,11}^{c}(t^{c}).
\end{align}
The qubit's state in the limit $n \rightarrow \infty$ is therefore given by
\begin{align}
\rho^\mu_S(t^h,t^c)=\rho_{11}^\mu(t^h,t^c)\ket{1}\bra{1}+\rho_{00}^\mu(t^h,t^c)\ket{0}\bra{0},
\end{align}
where
\begin{align}
\rho_{11}^\mu(t^h,t^c)\equiv& P^\mu(t^h,t^c)\rho_{0,11}^\mu(t^\mu)\notag\\
&+(1-P^\mu(t^h,t^c))\rho_{1,11}^\mu(t^\mu),\\
\rho_{00}^\mu(t^h,t^c)\equiv& P^\mu(t^h,t^c)\rho_{0,00}^\mu(t^\mu)\notag\\
&+(1-P^\mu(t^h,t^c))\rho_{1,00}^\mu(t^\mu).
\end{align}
We used the relations $\rho_{0,11}^\mu(t^\mu)+\rho_{0,00}^\mu(t^\mu)=1$ and $\rho_{1,11}^\mu(t^\mu)+\rho_{1,00}^\mu(t^\mu)=1$ for the algebra.
Note that $\rho_{00}^\mu(0,0)=P^\mu(0,0)$ and $\rho_{11}^\mu(0,0)=1-P^\mu(0,0)$.

\section{Work extraction as a measurement process}
\quad
The current prevailing view in quantum thermodynamics defines controllable and available energy as work and uncontrollable and unavailable energy as heat \cite{Vinjanampathy16}.
The uncontrollable energy changes, i.e., uncontrollable energy associated with changes of the state of the system in response to Hamiltonian changes and coupling with the environment, are often defined as heat.
The average heat for an infinitesimal time would be defined by
\begin{align}
\delta Q:=\mathrm{Tr}\left[d\rho_S \H_S\right],
\end{align}
where $\H_S$ is the Hamiltonian of the system.
On the other hand, since the time variation of the Hamiltonian of the system can be controlled, the energy changes associated with the change of the Hamiltonian are often defined as work \cite{Anders13}.
For example, in the standard model, which is often used by researchers in the community of statistical mechanics \cite{Talkner2007, Alicki1979, Kieu2006, Anders13, Tasaki00}, the dynamics of the composite system of the system and the bath(s) is assumed to be a unitary operation.
Under this assumption, the standard model defines the amount of the extracted work as
\begin{align}
\delta W:=\mathrm{Tr}[U\rho_{\mathrm{tot}} U^\dagger(\H+d\H)]-\mathrm{Tr}[\rho_{\mathrm{tot}} \H]
\label{standard_SB}
\end{align}
where $U$, $\rho_{\mathrm{tot}}$ and $\H$ are the unitary dynamics, the initial state, and the Hamiltonian of the total system.
When we consider an adiabatic work extraction in the standard model, we can omit the bath from eq.~\eqref{standard_SB}:
\begin{align}
\delta W:=\mathrm{Tr}[U_S\rho_S U^\dagger_S(\H_S+d\H_S)]-\mathrm{Tr}[\rho_S\H_S]
\label{infinitesimal_work}
\end{align}
where $U_S$ is the unitary dynamics of the system.
This standard model, however, is unsatisfactory in the sense that it does not specify how we measure the work $W$ of the closed system under the unitary dynamics, according to ref.~\cite{Hayashi17}.

\quad
To resolve the issue, Hayashi and Tajima \cite{Hayashi17} proposed a quantum scenario that specifies how a macroscopic system like us humans could quantify the work done by a quantum system \cite{Hayashi17}.
The quantum scenario considers a quantum mechanical work storage which exchanges work with the working substance \cite{Skrzypczyk14, Horodecki13}.
The energy transfer from the working substance to the work storage is described by the unitary time evolution of the whole system that consists of the system and the work storage.
The work-extraction process was modeled in analogy with the work-extraction process in classical thermodynamics \cite{Hayashi17}.
Consider a fuel cell as an example of classical thermodynamics.
Suppose that a set of microscopic systems that constitutes the fuel cell outputs energy to the macroscopic system and lifts a weight.
In this case, in order to quantify the work, a human being would measure the height of the weight lifted and quantify the work based on the results of the measurement.
By the same analogy, when the system of interest performs work to the work storage, the work performed is quantified by the quantum measurement of the work storage.

\quad
The work extraction process in the quantum scenario proposed by Hayashi and Tajima is called a completely positive work extraction model (CP-work extraction model).
In this indirect measurement process, four levels of the energy conservation law must be satisfied.
Let $\rho_I$ denote the initial state of the system, $w_j$ denote the work retrieved by the work storage system, and $p_j$ denote the probability of obtaining $w_j$.
Let $\{\mathcal{E}_j\}$ be the CP-instrument which indicates the change of the state of the system through the CP-work extraction process.
When the average value of $w_j$ is equal to the energy loss of the system, any state $\rho_S$ satisfies
\begin{align}
\mathrm{Tr}[\H_S\rho_S]=\sum_jw_j\mathrm{Tr}[\mathcal{E}_j(\rho_S)]+\sum_j\mathrm{Tr}[\H_S\mathcal{E}_j(\rho_S)],
\end{align}
which Hayashi and Tajima \cite{Hayashi17} called Level-1 energy conservation law.
Here, $\mathrm{Tr}[\H_S\rho_S]$ and $\sum_j\mathrm{Tr}[\H_S\mathcal{E}_j(\rho_S)]$ indicate the expectation values of the system's energy before and after the work extraction, respectively, while $\sum_jw_j\mathrm{Tr}[\mathcal{E}_j(\rho_S)]$ indicates the mean value of the extracted work.

\quad
When each measurement result $w_j$ is equal to the energy loss of the system, it is classified as Level-2, 3 or 4 energy conservation law depending on the presence or absence of measurement errors.
Since our model satisfies the Level-4 energy conservation law, which is the one with no measurement errors, let us focus on the Level-4 energy conservation here.

\quad
In order to explain the Level-4 energy conservation, let us introduce the initial state of the system $\Pi_x=\ket{h_x}\bra{h_x}$ using the energy eigenstate $\ket{h_x}$ of the Hamiltonian $\H_S$ associated with the eigenvalue $h_x$.
The extracted work $w_j$ is defined as the energy change in the Hamiltonian $\H_S$ as in eq.~(\ref{infinitesimal_work}).
Then the state after the work extraction which follows the Level-4 energy conservation law satisfies
\begin{align}
\mathcal{E}_j(\Pi_x)=P_{h_x-w_j}\mathcal{E}_j(\Pi_x)P_{h_x-w_j},
\end{align}
where $P_{h_x-w_j}$ is a projection operator associated with the energy eigenvalue $h_x-w_j$, which indicates the state after the work extraction.
In the CP-work extraction in the quantum scenarios, the work extraction process satisfies the Level-4 energy conservation law as long as the initial state of the work storage belongs to the eigenspace of its Hamiltonian $\H_W$ and the unitary $U$ and the Hamiltonian $\H$ on the whole system satisfies $[U,\;\H]=0$ \cite{Hayashi17}.
We will show later that our model satisfies the Level-4 energy conservation law.

\quad
The previous study \cite{Shirai21} of non-Markovian dynamics did not take into account macroscopic systems that exchange the work with the microscopic working substance.
We perform the work extraction processes using the quantum measurement theory \cite{Hayashi17} explained above so that we can describe extracting work consistently in the quantum Otto cycle.
By introducing a work storage as a new system, we describe the work extraction process as an indirect measurement model, in which we first let the qubit and the work storage interact with each other, and then measure the energy increase of the work storage.
In order to describe the variation of the qubit's Hamiltonian in the interaction process, we introduce a clock as an additional degree of freedom \cite{Horodecki13,Sagawa20}. The Hamiltonian of the total system $\tilde{\H}_{SE}$ contsists of the qubit, the clock and the work storage:
\begin{align}
\label{work_hamiltonian}
\tilde{\H}_{SE}=&\H_S^h\otimes\ket{0}\bra{0}_{C}\otimes I_W+\H_S^c\otimes\ket{1}\bra{1}_{C}\otimes I_W\notag\\
&+\tilde{\H}_W\:,
\end{align}
where the Hamiltonian of the work storage is given by
\begin{align}
&\tilde{\H}_W=I_S\otimes I_C\otimes(\omega^h-\omega^c)\ket{1}\bra{1}_W.
\end{align}
As the Hamiltonian of the system changes from $\H_S^h$ to $\H_S^c$, the clock changes from $\ket{0}\bra{0}_C$ to $\ket{1}\bra{1}_C$, and $\tilde{\H}_W$ is the energy extracted accordingly in the process.
This process describes the adiabatic expansion process in Fig.~\ref{otto_cycle}.
With respect to an adiabatic compression process, we replace $\H^h_S$ to $\H^c_S$, and $\H^c_S$ to $\H^h_S$ in eq.~(\ref{work_hamiltonian}).

\quad
More specifically, we first set the initial state of the whole system $\rho_{\mathrm{tot}}$ under the projection measurement to the qubit in order to extinguish the qubit's coherence,
\begin{align}
\label{work extraction_initial}
\rho^{\mu(0)}_{\mathrm{tot}}\equiv&\biggl(\rho^\mu_{11}(t^h,t^c)\ket{1}\bra{1}_S
+\rho^\mu_{00}(t^h,t^c)\ket{0}\bra{0}_S\biggr) \notag \\
&\;\otimes\ket{0}\bra{0}_C\otimes\ket{0}\bra{0}_W.
\end{align}
We then carry out the work extraction process by the following quenching unitary transformation, assuming that the operator acts on the state instantaneously:
\begin{align}
U&=\ket{010}\bra{000}+\ket{000}\bra{010}+\ket{111}\bra{100}\notag\\
&+\ket{100}\bra{111}+\ket{011}\bra{001}+\ket{001}\bra{011}\notag\\
&+\ket{110}\bra{110}+\ket{101}\bra{101},
\end{align}
where $\ket{ijk}\equiv\ket{i}_S\otimes\ket{j}_C\otimes\ket{k}_W\;(i,\;j,\;k=0,\;\mathrm{or}\;1)$.
This unitary operation commutes with the Hamiltonian $\tilde{\H}_{SE}$, and hence it conserves the total energy.
After the work extraction, the state turns into
\begin{align}
\label{work extraction_final}
\rho^{\mu(1)}_{\mathrm{tot}}&\equiv\rho^\mu_{11}(t^h,t^c)\ket{1}\bra{1}_S\otimes\ket{1}\bra{1}_C\otimes\ket{1}\bra{1}_W \notag \\
&+\rho^\mu_{00}(t^h,t^c)\ket{0}\bra{0}_S\otimes\ket{1}\bra{1}_C\otimes\ket{0}\bra{0}_W.
\end{align}
Since we assume the quenching unitary transformation, it does not take time to reach the final state $\rho_{\textrm{tot}}^{\mu(1)}$.
Upon turning the clock state from $\ket{0}\bra{0}_C$ in eq.~(\ref{work extraction_initial}) to $\ket{1}\bra{1}_C$ in eq.~(\ref{work extraction_final}), the work storage changes its state from $\ket{0}\bra{0}_W$ to $\ket{1}\bra{1}_W$ if the system is in the excited state $\ket{1}\bra{1}_S$, whereas it remains $\ket{0}\bra{0}_W$ if the system is in the ground state $\ket{0}\bra{0}_S$.

\quad
We finally extract the work from the work storage by performing the following projective measurement to eq.~(\ref{work extraction_final}):
\begin{align}
\label{projective_measurement}
P_0\equiv\ket{0}\bra{0}_W,\:\:\:P_1\equiv\ket{1}\bra{1}_W.
\end{align}
When the measurement result is associated with $\ket{1}\bra{1}_W$, it means that the work is extracted from the qubit to the work storage. 
The probability of getting the measurement result is $\rho^\mu_{11}(t^h,t^c)$.
The amount of the extracted work is derived as the energy flow into the work storage;
\begin{align}
&\mathrm{Tr}[\tilde{\H}_W(I_S\otimes I_C\otimes\ket{1}\bra{1}_W-I_S\otimes I_C\otimes\ket{0}\bra{0}_W))]\\
&=\omega^h-\omega^c.
\end{align}
Meanwhile, the probability of getting the measurement result associated with $\ket{0}\bra{0}_W$ is $\rho^\mu_{00}(t^h,t^c)$.
In this case, the amount of the extracted work to the work storage is $0$.
Therefore, the expectation value of the harvested work under measurement probability is amounted to
\begin{align}
\langle dW \rangle=&\rho^\mu_{11}(t^h,t^c)\;\mathrm{Tr}\biggl[\tilde{\H}_W\bigl(I_S\otimes I_C\otimes\ket{1}\bra{1}_W\notag\\
&\:\:\:\:\:\:\:\:\:\:\:\:\:\:\:\:\:\:\:\:\:\:\:\:\:\:\:\:\:\:\:-I_S\otimes I_C\otimes\ket{0}\bra{0}_W\bigr)\biggr]\\
=&\rho^\mu_{11}(t^h,t^c)\left(\omega^h-\omega^c\right)\label{work_definition}
\end{align}
After all, this result is the same as in ref.~\cite{Shirai21}, which does not take into account macroscopic systems that exchange the work with the microscopic working substance.

\quad
Let us note that if the qubit's density matrix has off-diagonal elements in eq.~(\ref{work extraction_initial}), the accuracy of detecting the amount of extracted work~(\ref{work_definition}) by the measurement result decreases, because of a tradeoff between the coherence after the work-extraction and the accuracy of the measurement of the extracted work \cite{Hayashi17} (see also Theorem 1 in the supplemental material of Ref.~\cite{Tajima2018} and Lemma 3 in Ref.\cite{Tajima2020}); in short, the decoherence enhances the accuracy.
As a countermeasure, we eliminate the off-diagonal elements of the qubit's state by the projection measurement, and hence this problem does not occur \cite{Hayashi17}.
Because of the assumption, the estimate of the extracted work~(\ref{work_definition}) is equal to the previous one \cite{Shirai21}; in other words, the present formulation gives a precise physical background to the estimate~(\ref{work_definition}).
Let us finally note that the present work extraction process is applicable to general models without the projection measurement that eliminates the off-diagonal elements.

\section{Results}
\quad
We examine the interaction energy in detail and reveal that the non-Markovianity appears mainly in three ways depending on the interaction time between the qubit and the bath.
In order to clarify these effects, we propose an index which indicates the magnitude of the non-Markovianity.
Because of the non-Markovian effects, we can make the quantum Otto cycle under non-Markovian dynamics behave differently from the conventional quantum heat engine by controlling the interaction time between the qubit and the bath.\\
\quad
Note that all the results here are for the limit cycle that we introduced at the end of Sec.~\ref{Model}, which is achieved after convergence.
The interaction time $t^\mu$ is a parameter fixed through all the cycles; it should not be considered as a parameter that is controllable during a cycle.
In all the numerical results, we fix $\lambda^h=\lambda^c=0.01$ and $\Omega^h=\Omega^c=0.4$ with $k_B=\hbar=1$.

\subsection{Non-Markovianity of the interaction energy}
\label{Non-Markovianity of the interaction energy}
\quad
Let us analyze the non-Markovian dynamics of the energy expectation value in the two isochoric processes. The energy changes of the qubit $\Delta E_S^{\mu}(t^h,t^c)$, the bath $\Delta E_B^{\mu}(t^h,t^c)$, and the interaction energy $\Delta E_I^{\mu}(t^h,t^c)$ are given by
\begin{align}
\Delta E_S^{\mu}(t^h,t^c)&=\langle \H_S^{\mu}(t^\mu)\rangle_S - \langle \H_S^{\mu}(0)\rangle_S,\\
\Delta E_B^{\mu}(t^h,t^c)&=\langle \H_B^{\mu}(t^\mu)\rangle_B - \langle \H_B^{\mu}(0)\rangle_B, \label{bath energy}\\
\Delta E_I^{\mu}(t^h,t^c)&=-\Delta E_S^{\mu}(t^h,t^c)-\Delta E_B^{\mu}(t^h,t^c), \label{Interaction energy}
\end{align}
where $\H^\mu(t^\mu)$ is the Hamiltonian after the qubit interacts with the $\mu$th bath over time $t^\mu$ and $\langle \cdot \rangle_j\equiv \tr[\:\cdot \:\rho_j]$ with $j=S,B$.
We define the change of the interaction energy as in eq.~(\ref{Interaction energy}) so that the total energy may be conserved.
The energy changes are derived as follows by taking the interaction time as a parameter \cite{Shirai21};
\begin{align}
\Delta E_S^{\mu}(t^h,t^c)&=\omega^{\mu} \left( \rho^\mu_{11}(t^h,t^c) - \rho^\mu_{11}(0,0)\right), \label{NM_ES}\\
\Delta E_B^{\mu}(t^h,t^c)&=\omega^{\mu} \left( \rho^\mu_{11}(t^h,t^c) - \rho^\mu_{11}(0,0)\right) \notag\\
&\; +\int_0^{t^\mu}d\tau \Biggl[ \biggl\{2[P^\mu(t^h,t^c)\rho_{0,00}^\mu(\tau)\notag\\
&\; +(1-P^\mu(t^h,t^c))\rho_{1,00}^\mu(\tau)]-1\biggl\}D_1(\tau)\mathrm{sin}(\omega^\mu \tau) \notag\\
&\;+D_2(\tau)\mathrm{cos}(\omega^\mu \tau)\Biggr], \label{NM_EB}\\
\Delta E_I^{\mu}(t^h,t^c)&=-\int_0^{t^\mu}d\tau \Biggl[ \biggl\{2[P^\mu(t^h,t^c)\rho_{0,00}^\mu(\tau)\notag\\
&\; +(1-P^\mu(t^h,t^c))\rho_{1,00}^\mu(\tau)]-1\biggl\}D_1(\tau)\mathrm{sin}(\omega^\mu \tau)\notag\\
&\;+D_2(\tau)\mathrm{cos}(\omega^\mu \tau)\Biggr], \label{NM_EI}
\end{align}
where $P^\mu(t^h,t^c)$ is a constant defined in eq.~(\ref{large_p}).
The energy change of the bath in eq.~(\ref{NM_EB}) is derived by using the full counting statistics \cite{Guarnieri16, Shirai21}, which we review in Appendix.~\ref{Full Counting Statistics}.
For the Markovian dynamics, we find that for any $t^\mu$, the second term of eq.~(\ref{NM_EB}) vanishes and $\Delta E_I^{\mu}=0$, and hence $\Delta E_S^{\mu}=\Delta E_B^{\mu}=\omega^{\mu} \left( \rho^{M,\mu}_{11}(t^h,t^c) - \rho^{M,\mu}_{11}(0,0)\right)$; see Appendix.~\ref{Markovian dynamics}.

\quad
We reveal the following three distinct properties of the non-Markovian dynamics.

\begin{figure}
\includegraphics[width=0.5\textwidth]{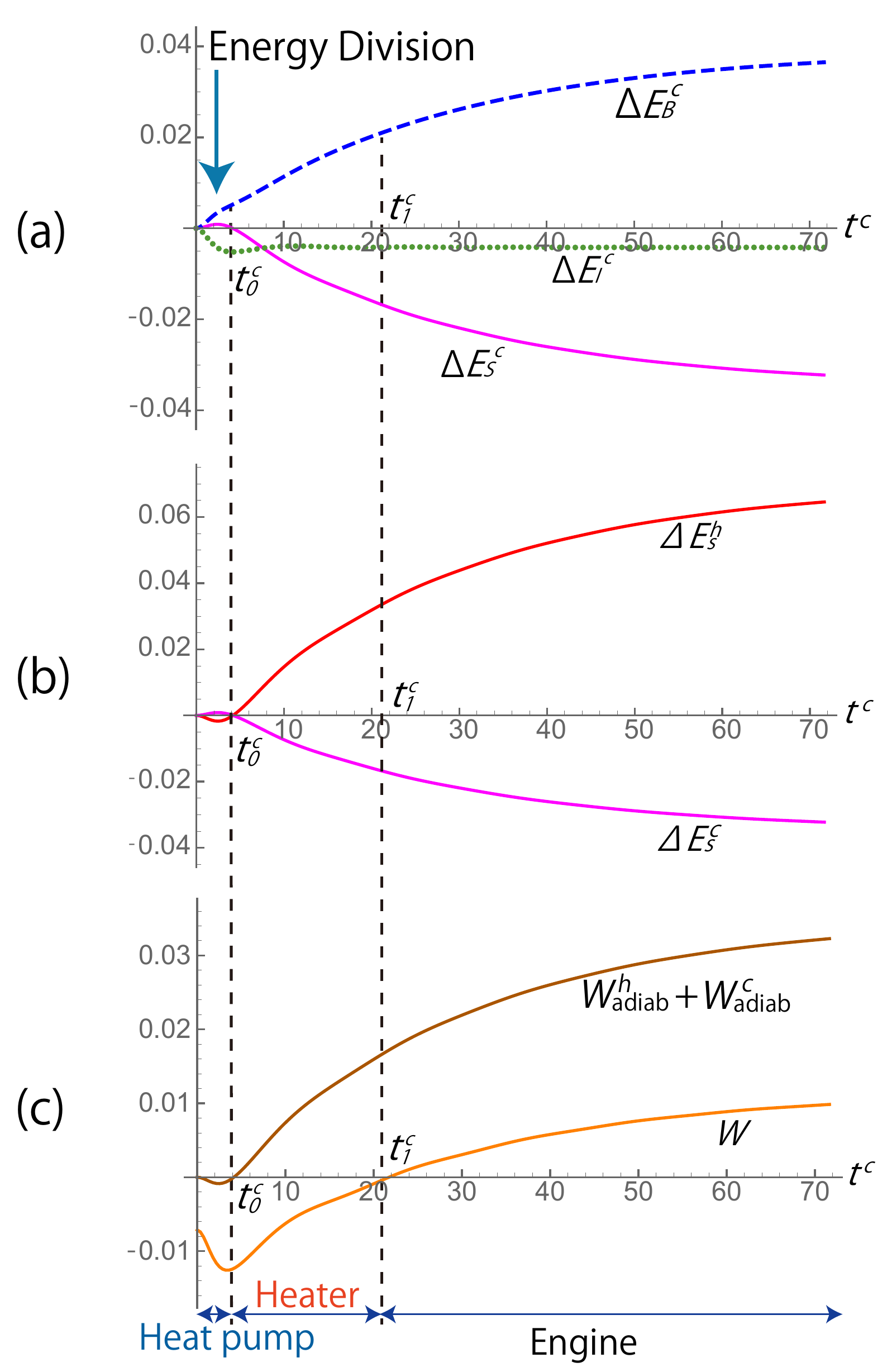}
\caption{(a) Non-Markovian dynamics of the change of the qubit energy, $\Delta E_S^c$, of the cold-bath energy, $\Delta E_B^c$, and of the interaction energy, $\Delta E_I^c$, depending on the interaction time between the qubit and the cold bath, $t^c$. In the regime of the short interaction time ($0\leq t^c\leq t^c_0$), the energy division occures, whereas the interaction energy remains negative in the regime of the long interaction time.
(b) $\Delta E_S^h$ is the qubit's energy change during the time it interacts with the hot bath and $\Delta E_S^c$ is the one with the cold bath. At $t^c=t_0$, the sign of each qubit's energy changes.
(c) The total work $W$ and the work extracted in the adiabatic process $W_\mathrm{adiab}^h+W_\mathrm{adiab}^c$. At $t^c=t_0^c$, the sign of the adiabatic work changes and at $t^c=t_1^c$, the one of the total work changes.
The parameters are set to $t^h=60$, $\omega^c/\omega^h=0.5$, and $T^c/T^h=0.2$ in (a), (b) and (c).}
\label{EnergyChange_omega_050_T_020}
\end{figure}

\quad
\textit{Energy division in the region of the short interaction time}:
Figure~\ref{EnergyChange_omega_050_T_020} shows the energy changes during the isochoric process in which the qubit interacts with the cold bath for time $t^c$. 
Counterintuitively, the energy change of the qubit is positive in the region of the short interaction time ($0<t^c<t_0$), which means that the qubit absorbs energy during the interaction with the cold bath for a short time.
We presume that this is because the qubit and the bath devide the slight decrease of the interaction energy $\Delta E_I^\mu$; we call this phenomenon ``energy division''.
The energy division is a phenomenon slightly different from the so-called energy backflow \cite{Guarnieri16}, in which the energy of the system of interest that flows into the bath comes back to the system during the time evolution.

\begin{figure}
\includegraphics[width=0.45\textwidth]{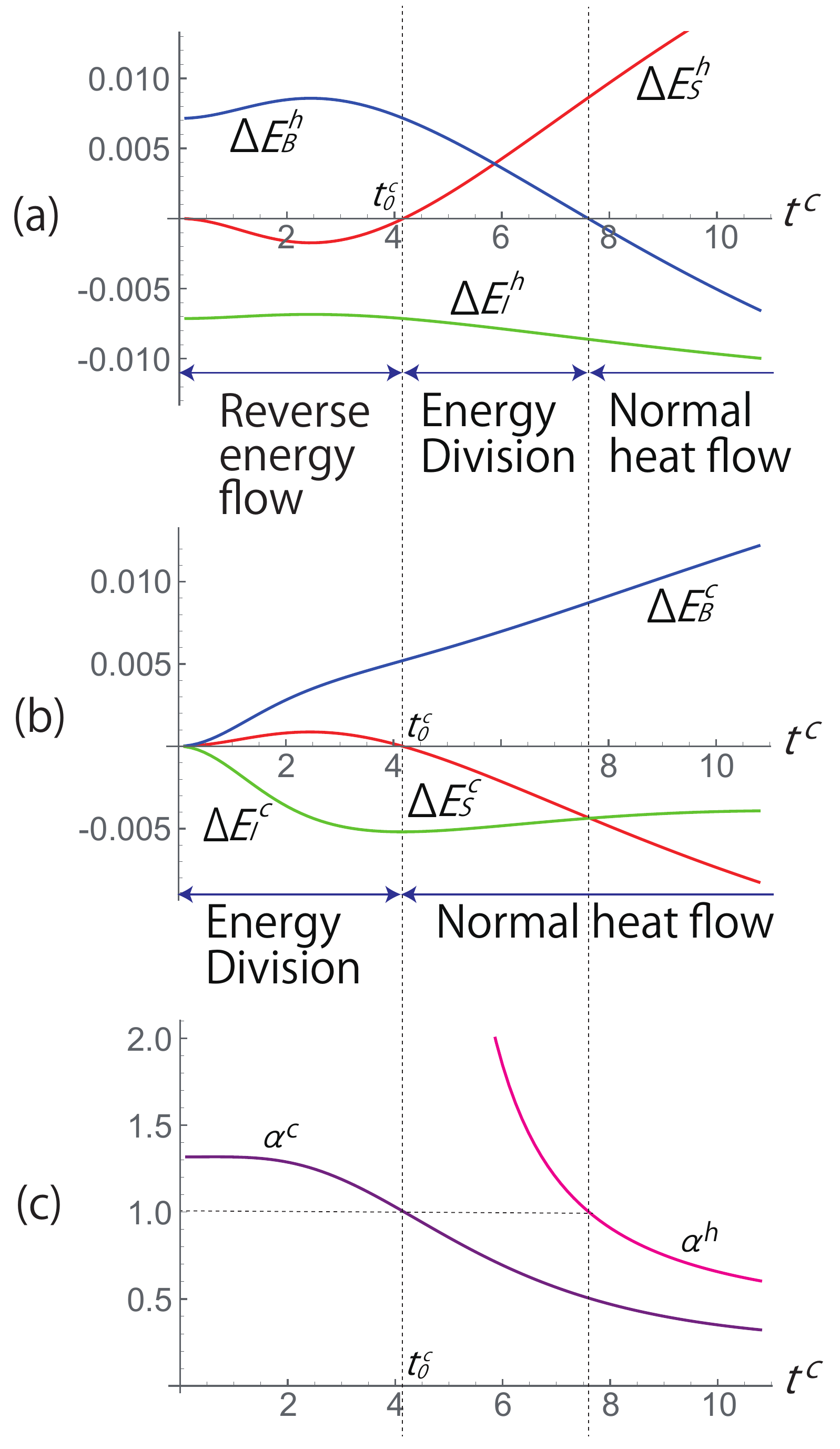}
\caption{(a) The reverse energy flow, the energy division and the normal heat flow between the qubit and the hot bath.
The qubit's energy is denoted by $\Delta E_S^h$, the hot bath's energy by $\Delta E_B^h$, and the interaction energy by $\Delta E_I^h$.
(b) The energy division and the normal energy flow between the qubit and the cold bath.
The qubit's energy is denoted by $\Delta E_S^c$, the cold bath's energy by $\Delta E_B^c$, and the interaction energy by $\Delta E_I^c$.
(c) The indexes of the non-Markovianity in terms of the energy changes $\alpha^c$ and $\alpha^h$.
The parameters are set to $t^h=60$, $\omega^c/\omega^h=0.5$ and $T^c/T^h=0.2$.}
\label{EnergyReversal_hot}
\end{figure}

\quad
\textit{Reverse energy flow in the region of the short interaction time}:
We also find as shown in Fig.~\ref{EnergyReversal_hot} (a) that the signs of the energies of the qubit and the hot bath are reversed compared to the normal energy flow.
This behavior only happens when the qubit interacts with the hot bath; we numerically confirm that this behavior does not appear when the qubit interacts with the cold bath as shown in Fig.~\ref{EnergyReversal_hot} (b).

\quad
\textit{Non-Markovian behavior of the energy in the region of the long interaction time}:
In all cases that we have numerically analyzed, the interaction energy $\Delta E_I^\mu$ is negative for any interaction time $t^\mu$, which means that there is an attractive interaction between the qubit and the bath. 
Therefore, we have to add an extra work to detach the qubit from the bath in order to go to the next adiabatic process of the cycle.
We should count this detachment work $W_\textrm{detach}^{\mu}=\Delta E_I^\mu\;(<0)$ into the total work of the cycle together with the work extracted in the adiabatic process $W_\textrm{adiab}^h=\left(\omega^h-\omega^c\right)\rho^h_{11}(t^h,t^c)$ and $W_\textrm{adiab}^c=\left(\omega^c-\omega^h\right)\rho^c_{11}(t^h,t^c)$.
We thereby obtain the total work in the form
\begin{align}
\label{work}
W=W_{\textrm{adiab}}^h+W_{\textrm{adiab}}^c+W_{\textrm{detach}}^h+W_{\textrm{detach}}^c,
\end{align}
where the work in the direction from the working substance to the outside is set to be positive, while $W_\textrm{detach}^{\mu}$ is always negative, which indicates that this is the work done from the outside.
We have numerically confirmed that the integrand of the expectation value of the interaction energy, $\Delta E_I^\mu(t^h,t^c)$ in eq.~(\ref{NM_EI}), vanishes in the long interaction time limit $t^\mu \rightarrow \infty$.
Therefore, the negative interaction energy converges to a constant value as shown in Fig.~\ref{EnergyChange_omega_050_T_020} (a).

\quad
\textit{The classification of Non-Markovianity in terms of energy}:
Let us classify the counterintuitive behavior based on the features of the reverse energy flow, the energy division, and the normal energy flow; see Table~\ref{table:Classification} (a).
The normal energy flow is defined as a heat flow from the hot bath to the qubit and a heat flow from the qubit to the cold bath.
Note that this energy flow includes the energy flow from the interaction energy to the bath or to the qubit.

\quad
In order to describe the peculiar behavior in the non-Markovian dynamics, we propose an index of the non-Markovianity in terms of the energy changes.
We define the index as the ratio of the interaction energy to the bath's energy and to the qubit's energy:
\begin{align}
\label{NMpara}
\alpha^c\equiv\left|\frac{\Delta E^c_I}{\Delta E^c_B}\right|
\:\:\mathrm{and}\:\:
\alpha^h\equiv\left|\frac{\Delta E^h_I}{\Delta E^c_S}\right|,
\end{align}
where the superscripts $c$ and $h$ indicate the energy changes after the qubit interacts with the cold bath and with the hot bath, respectively.
The occurrence of the energy division $|\Delta E^c_I|=|\Delta E^c_S|+|\Delta E^c_B|$ yields $\alpha^c\geq 1$.
For $1/2\leq\alpha^c\leq1$, the interaction energy is a dominant contributor to the bath's energy, $\Delta E^c_S\leq\Delta E^c_I$, while for $\alpha^c\leq1/2$, the qubit's energy is a dominant one, $\Delta E^c_I\leq\Delta E^c_S$.
In addition, we analytically confirm that the energy division between the qubit and the cold bath and the reverse energy between the qubit and the hot bath occur in the same region of the interaction time $t^h$ and $t^c$.
In the same region of $\alpha^c\geq 1$, the occurrence of the reverse energy flow occurs between the qubit and the hot bath which we described in Fig.~\ref{EnergyChange_omega_050_T_020} (b).
In the region $\alpha^c\leq1$ and $\alpha^h\geq1$, the energy division occurs between the qubit and the hot bath.
The normal heat flow occurs from the hot bath to the qubit in the region $\alpha^c\leq1$ and $\alpha^h\leq1$.

\begin{figure}
\includegraphics[width=0.5\textwidth]{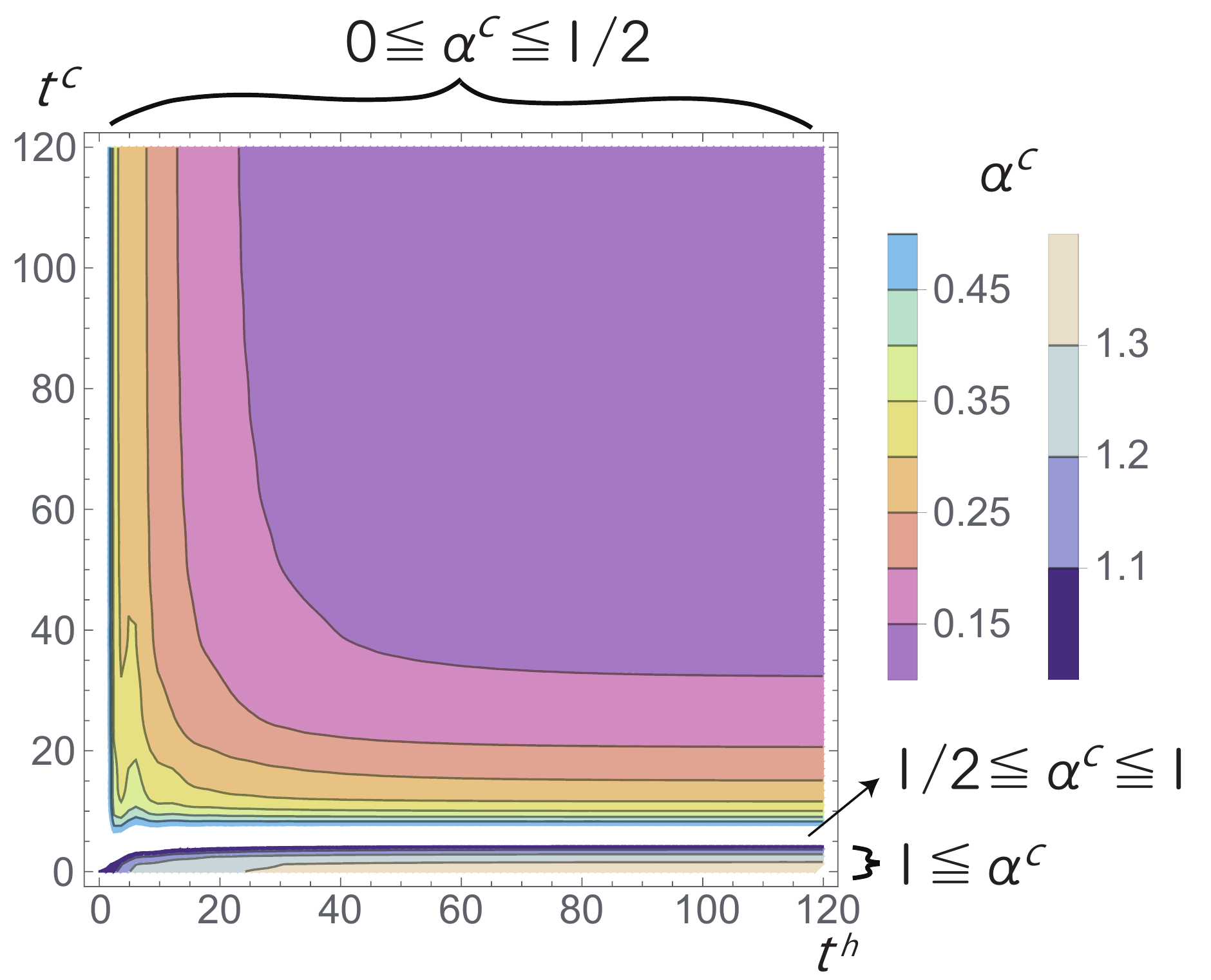}
\caption{A contour plot of the non-Markovian index $\alpha^c$. The energy division occurs in the short-interaction-time region of $t^h$ and $t^c$. The longer the interaction times $t^h$ and $t^c$, the closer $\alpha^c$ gets to zero, which means that the Markov approximation becomes valid. For simplicity, we make the contour plot in the region $\alpha^c\geq 1$ and $\alpha^c\leq 1/2$ individually, while the white region indeicates $1/2 \leq \alpha^c \leq 1$. The parameters are set to $\omega^c/\omega^h=0.5$ and $T^c/T^h=0.2$.}
\label{EnergyDivisionContour}
\end{figure}

\quad
Figure~\ref{EnergyReversal_hot} exemplifies the $t^c$ dependence of the heat flow and the non-Markovianity index (\ref{NMpara}).
This numerical calculation shows that the negative interaction energy is still a dominant contributor to the function of the quantum heat engine in the region of the long interaction time. Dependence of the non-Markovian index $\alpha^c$ on the interaction time with the hot bath $t^h$ and the one with the cold bath $t^c$ is shown in Fig.~\ref{EnergyDivisionContour}.
We also note that this index becomes zero in the Markovian dynamics because the expectation value of the interaction energy is always zero.

\begin{table}[h]
 \caption{Classification of the behavior of Non-Markovian dynamics when the qubit interacts with (a) the hot bath and (b) the cold bath.}
 \label{table:Classification}
\vskip\baselineskip
 \centering
  \begin{tabular}{ccc}
   \hline
    (a)& $\Delta E_B^h>0$ & $\Delta E_B^h<0$ \\
   \hline \hline
   $\Delta E_S^h>0$ & Energy Division & Normal Energy Flow \\
   $\Delta E_S^h<0$ & $\:$Reverse Energy Flow$\:$ & $-$ \\
   \hline
  \end{tabular}
\vskip\baselineskip
  \begin{tabular}{ccc}
   \hline
    (b)& $\Delta E_B^c>0$ & $\Delta E_B^c<0$ \\
   \hline \hline
   $\Delta E_S^c>0$ & Energy Division & $\:$Reverse Energy Flow $\:$ \\
    & & (No Occurrence) \\
   $\Delta E_S^c<0$ & $\:$Normal Energy Flow$\:$ & $-$ \\
   \hline
  \end{tabular}
\end{table}

\subsection{Heat pump and Heater}
\label{Heat pump and Heater}
\quad
A ``heat pump'' and a ``heater'' appear depending on the interaction time. We call the system a ``heat pump'' when the qubit acquires the work $W$ from outside and transfers heat from the cold bath to the hot bath.
On the other hand, we call the system a ``heater'' when the qubit transfers heat oppositely to the heat pump based on the work it acquires from outside.

\quad
More specifically, we have a ``heat pump'' for $0<t^c<t_0^c$ and a ``heater'' for $t_0^c<t^c<t_1^c$ as is exemplified in Fig.~\ref{EnergyChange_omega_050_T_020} (c).
In the former ``heat pump'' region $0<t^c<t_0^c$, the energy division occurs as is indicated in Fig.~\ref{EnergyChange_omega_050_T_020} (a).
The qubit itself loses its energy through one cycle; the absolute value of the negative value of $\Delta E_S^h$ exceeds the positive value of $\Delta E_S^c$ as shown in Fig.~\ref{EnergyChange_omega_050_T_020} (b).
This is because the probability of the excited state $\rho_{11}^c(t^h,t^c)$ of the qubit after interacting with the cold bath exceeds the probability after interacting with the hot bath $\rho_{11}^h(t^h,t^c)$ in the region $0< t^c < t_0^c$.

\quad
In the same region, not only the total work is negative but the work done through the adiabatic processes, that is, $W^h_{\mathrm{adiab}}+W^c_{\mathrm{adiab}}=(\omega^h-\omega^c)\rho^h_{11}(t^h,t^c)+(\omega^c-\omega^h)\rho^c_{11}(t^h,t^c)$, is negative because the probability of the excited state $\rho_{11}^c(t^h,t^c)$ exceeds the probability $\rho_{11}^h(t^h,t^c)$.
In this sense, the qubit is cooled through this cycle under a work from outside, which function we refer to as the heat pump.
This property could be used for cooling the qubit in quantum computers.

\quad
We analytically find that $\Delta E_S^h$ and $\Delta E_S^c$ vanish at the same value of $t^c=t_0^c$, beyond which the heat absorption and extraction processes are switched.
In the ``heater'' region $t_0^c<t^c<t_1^c$, the qubit acquires heat from the hot bath and discard heat to the cold bath (Fig.\ref{EnergyChange_omega_050_T_020} (b)).
Although the work done through the adiabatic processes turns to positive (Fig.~\ref{EnergyChange_omega_050_T_020} (c)), the total work is still negative in the region $t_0^c<t^c<t_1^c$, because the absolute value of the negative detachment work $W^h_{\mathrm{detach}}+W^c_{\mathrm{detach}}$ exceeds the work in the adiabatic processes $W^h_{\mathrm{adiab}}+W^c_{\mathrm{adiab}}$ (Fig.~\ref{EnergyChange_omega_050_T_020} (c)).
In this sense, the cycle transfers heat from the hot bath to the cold bath based on the work from the outside, and hence we refer to this function as the heater.

\quad
Beyond the point $t^c=t_1^c$, the total work finally turns to positive.
The cycle covers heat to the work as a normal engine.
We note that in the Markovian dynamics, the heat-pump and heater regions do not appear; in all regions, the quantum Otto cycle functions as an engine as shown in Appendix \ref{Markovian dynamics}.

\quad
Figure \ref{nonMarkov_omega_050_T_020} (a)-(c) show the work extraction and the heat absorption for both the interaction times $t^h$ and $t^c$.
In order to characterize the regions of the heat pump and the heater, we use the coefficient of performance (COP) \cite{Riffe94}:
\begin{align}
\label{COP}
\mathrm{COP}\equiv\frac{|\Delta E_S^c|}{|W|},
\end{align}
where $|\Delta E_S^c|$ is the amount of heat that the qubit absorbs from the cold bath and $W$ is the total work (\ref{work}).
On the other hand, the efficiency of this cycle, when it works as an engine, is given as follows:
\begin{align}
\label{efficiency}
\eta \equiv\frac{W}{\Delta E_S^h},
\end{align}
where $\Delta E_S^h$ is the heat that the qubit absorbs from the hot bath.
We plot them together in Fig.~\ref{nonMarkov_omega_050_T_020} (d).

\quad
Incidentally, the heat pump occurs only in the domain of the short interaction time of $t^c$ as shown in Fig.~\ref{nonMarkov_omega_050_T_020} (d).
This assymmetry appears because the interaction energy is always negative no matter if the working substance interacts with either hot or cold bath.
For example, the energy division always divides the negative interaction energy to a positive system energy and a positive bath energy.
It would be normal for the qubit's energy change to be positive during the interaction with the hot bath, even without the energy division. 
However, it would be abnormal for the qubit's energy change to be positive when interacting with a cold bath; it only happens because of the energy division as exemplified in Fig.~\ref{EnergyChange_omega_050_T_020}.
This causes the assymmetric appearance of the heat pump.
As we wrote above, we numerically confirmed that the interaction energy is negative as far as we analyzed, but the analytical confirmation is still an open question.

\begin{figure*}[t]
\includegraphics[width=1\textwidth]{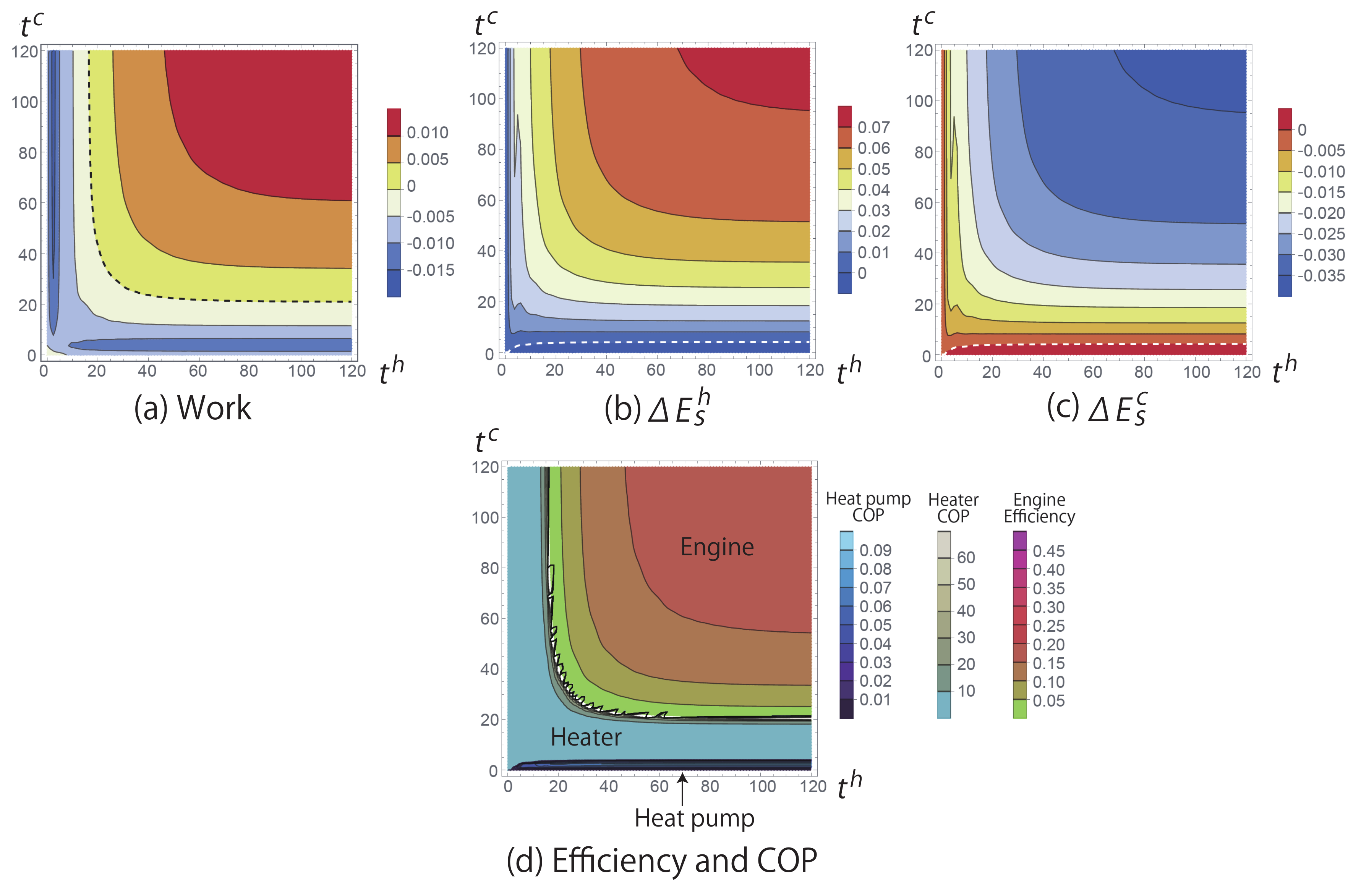}
\caption{Work, energy expectation values $\Delta E_S^h$ and $\Delta E_S^c$, efficiency $\eta$ and COP when $\omega_c/\omega_h=0.5$, $T_c/T_h=0.2$ for the interaction times $t^c$ and $t^h$. Each of the dashed lines in the top panels indicates the boundary where the sign flips.
Figure \ref{EnergyChange_omega_050_T_020} corresponds to the cross section on the vertical line $t^h=60$.}
\label{nonMarkov_omega_050_T_020}
\end{figure*}

\begin{figure*}[ht]
\begin{center}
\includegraphics[width=1\textwidth]{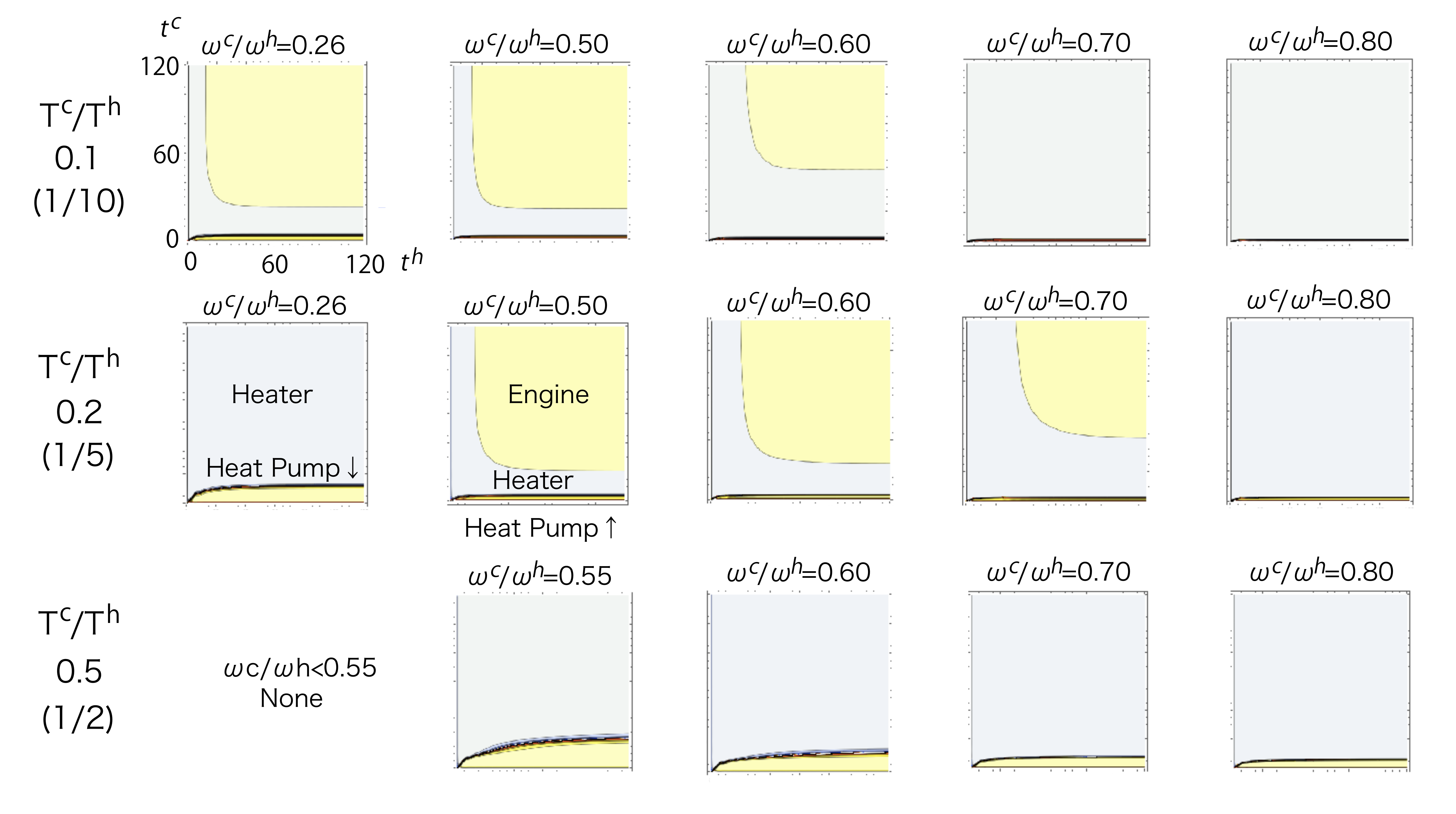}
\caption{The dependency of the function of the cycle on $T^c/T^h$ and $\omega^c/\omega^h$ in the ranges $0\leq t^h \leq 120$ and $0\leq t^c \leq 120$ on the horizontal and vertical axes, respectively.}
\label{nonMarkov_Temp_Sensitivity}
\end{center}
\end{figure*}

\quad
We also find the phase diagram of the heat pump, heater and the engine for various values of $T^c/T^h$ and $\omega^c/\omega^h$, as in Fig.~\ref{nonMarkov_Temp_Sensitivity}.
The heat pump and the heater appear in Fig.~\ref{nonMarkov_Temp_Sensitivity} only for $T^c/T^h\ll\omega^c/\omega^h$ on the right side and for $T^c/T^h\approx\omega^c/\omega^h$ on the bottom left side.
For $T^c/T^h\ll\omega^c/\omega^h$, we have $\eta_O\ll\eta_C$, where $\eta_O=1-\omega^c/\omega^h$ is the efficiency of the Otto cycle and $\eta_C=1-T^c/T^h$ is the Carnot efficiency.
This means that when the quantum Otto cycle is inefficient, it loses its function as an engine and acquires another function instead.
In the region $T^c/T^h\approx\omega^c/\omega^h$, the closer $\omega^c/\omega^h$ becomes to $T^c/T^h$, the closer $\eta_O$ becomes to $\eta_C$.
Before $\eta_O$ reaches $\eta_C$, there appears limitation to functioning as the engine of the non-Markovian quantum Otto cycle.
We note that the previous study \cite{Shirai21} analyzed the behavior of the quantum Otto cycle in the parameter region in which the effective temperature of the working substance exceeds the temperature of the hot bath, so that the efficiency of the quantum Otto cycle exceeds the Carnot efficiency, and hence there is no positive work harvesting.

\subsection{Alternative function from the viewpoint of the energy expectation value of the bath}
\begin{figure*}[h]
\includegraphics[width=0.8\textwidth]{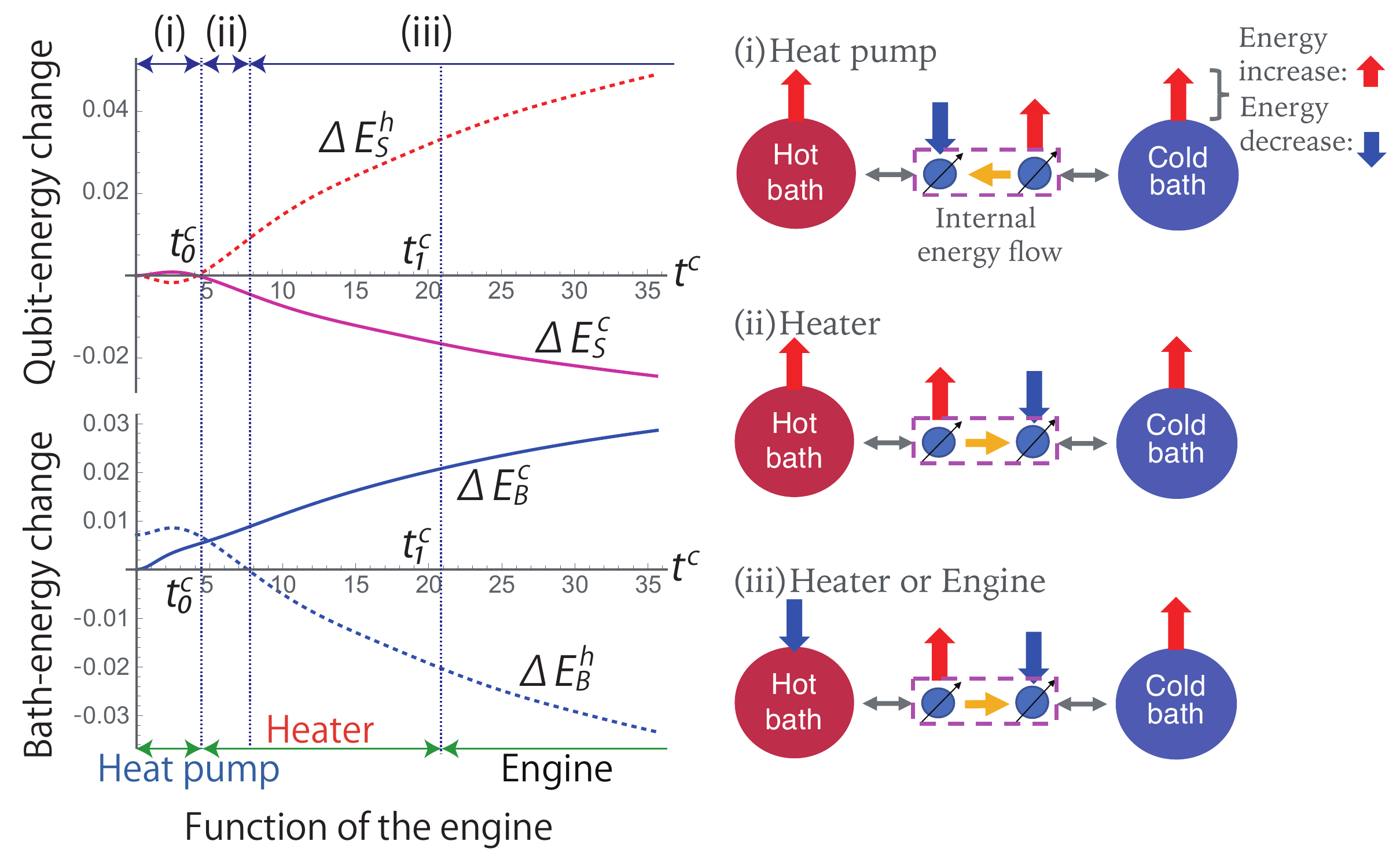}
\caption{The energy changes of the bath and the qubit, where $\Delta E_S^\mu$ ($\mu$=h or c) is the qubit's energy change when it interacts with the hot or cold bath, respectively, and $\Delta E_B^\mu$ ($\mu$=h or c) is the energy change of the hot or cold bath, respectively. The axis $t^c$ indicates the interaction time with the cold bath. The parameters are set to $t^h=60$, $\omega^c/\omega^h=0.5$, and $T^c/T^h=0.2$.}
\label{BathEnergy}
\end{figure*}

There is a counterintuitive energy change when we focus on the energy expectation value of the bath as shown in Fig.~\ref{BathEnergy}.
In the heat-pump region (i) in Fig.~\ref{BathEnergy}, the qubit itself pumps heat from the cold bath to the hot bath.
Nevertheless, the energy expectation values of the hot and cold baths increase.
This is because the negative interaction energy is divided into the qubit and the cold bath, which we called the energy division in Sec.~\ref{Non-Markovianity of the interaction energy}, and because the energy flows from the qubit to the hot bath, which we called the reverse energy flow in Sec.~\ref{Non-Markovianity of the interaction energy}.
In a part of the heater region (the region (ii) in Fig.~\ref{BathEnergy}), the energy expectation values of the hot and cold baths still increase, which is because of the energy division of the qubit and the hot bath.
In the other part of the heater region and the engine region (the region (iii) in Fig.~\ref{BathEnergy}), the energy expectation value of the cold bath increases while the one of the hot bath decreases.
We numerically confirmed that the expectation value of the cold bath does not decrease (Table~\ref{table:Classification} (b)).

\section{Conclusions and perspectives}
\quad
In this paper, we focused on the quantum-thermodynamic problem of theoretical description of the interactions between macroscopic systems and microscopic quantum ones.
We explored the energy flow of the quantum Otto cycle in the non-Markovian processes, examining heat absorption and emission in isochoric processes as well as work extraction in adiabatic processes.
As a result, we revealed a counterintuitive feature of the non-Markovian quantum Otto cycle; we found that the cycle functions as an engine, a heat pump and a heater, depending on the interaction time with the baths.\\
\quad
The non-Markovianity mainly has three effects on the cycle depending on the interaction time: the energy division as well as the reverse energy flow from the qubit to the hot bath occur in the region of the short interaction time; the interaction energy remains negative in the region of the long interaction time.
We quantified these effects by defining an index of non-Markovianity in terms of the energy division of the interaction energy.
We also focused on the energy expectation value of the hot and cold baths, revealing that the heat flow from the cold bath to the hot bath does not occur despite of the work.
The qubit itself loses its energy if we shorten the interaction time between the qubit and the bath. In this sense, the qubit is cooled through the cycle, which property could be used for cooling the qubits in quantum computers.
This non-trivial behavior does not appear in the Markovian dynamics.\\
\quad
In the isochoric processes, the interaction energy between the qubit and the bath becomes negative.
In order to complete the cycle, we have to add an extra work to detach the qubit from the bath.
When we count this detachment work into the total work $W$, the work becomes negative depending on the interaction time.
We also showed that this non-trivial energy flow and the negative work make the quantum Otto cycle behave as a heat pump and a heater.\\
\quad
We also discussed the work extraction from a microscopic system to a macroscopic system like us humans \cite{Hayashi17}.
We modeled the work extraction processes using quantum measurements explicitly.\\
\quad
An open question is how to treat the feedback based on the quantum measurement information after the detachment of the qubit and the bath.
We will also investigate a rigorous trade-off relationship between the efficiency and the power including the negative interaction energy \cite{PhysRevE.96.022138}.

\section*{Acknowledgements}
One of the authors, M.I., is supported by Quantum Science and Technology Fellowship Program (Q-STEP), the University of Tokyo.
N.H.'s work is partly supported by JSPS KAKENHI Grant Numbers JP19H00658, JP21H01005 and JP22H01140.
H.T.'s work is supported by JSPS Grants-in-Aid for Scientific Research No. JP19K14610, No. JP22H05250, JST PRESTO No. JPMJPR2014, and JST MOONSHOT No. JPMJMS2061.

\appendix

\section{Full Counting Statistics}
\label{Full Counting Statistics}
We use the method of full counting statistics \cite{Guarnieri16, Shirai21, RevModPhys.81.1665} for evaluation of the physical quantities of the bath.
Let us review it here to make our description self-contained.
In this method, the concept of two-point measurement for the bath is used.
For a measurable physical quantity $Q$ of the environmental system, the measurement operator is given by $\{\hat{\Pi}_q=\ket{q}\bra{q}\}$, using the eigenstate $\{\ket{q}\}$ of $Q$.
For the two-point measurement, projective measurements $\hat{\Pi}_{q_0}=\ket{q_0}\bra{q_0}$ and $\hat{\Pi}_{q_t}=\ket{q_t}\bra{q_t}$ are performed on the environmental system at time $0$ and $t$, respectively.
Hence, the probability of obtaining the measurement results $q_0$ and $q_t$ at time $0$ and $t$, respectively, is
\begin{align}
P(q_t,q_0)=&\mathrm{Tr_{SB}}[\;(I\otimes\hat{\Pi}_{q_t})\;U(t,0)\;(I\otimes\hat{\Pi}_{q_0})\notag\\
&\rho_{\mathrm{tot}}(0)\;(I\otimes\hat{\Pi}_{q_0})\;U^\dagger(t,0)\;(I\otimes\hat{\Pi}_{q_t})].
\end{align}
Therefore, the probability distribution of the difference $\Delta q=q_t-q_0$ between the two measurement results $q_t$ and $q_0$ is
\begin{align}
P_t(\Delta q)=\sum_{q_0,q_t}\delta(\Delta q-(q_t-q_0))P(q_t,q_0).
\end{align}

\quad
For the calculation of the expectation value of physical quantities, it is useful to define the cumulant generating function for $\Delta q$:
\begin{align}
&S_t(\chi)=\ln\int_{-\infty}^\infty d(\Delta q)P_t(\Delta q)e^{i\chi\Delta q}\notag\\
&=\ln\int_{-\infty}^\infty d(\Delta q)\sum_{q_0,q_t}\delta(\Delta q-(q_t-q_0))P(q_t,q_0)e^{i\chi\Delta q}\notag\\
&=\ln\Biggl[\sum_{q_0,q_t} \mathrm{Tr_{SB}}[\;(I\otimes\hat{\Pi}_{q_t})\;U(t,0)\;(I\otimes\hat{\Pi}_{q_0})\;\rho_{\mathrm{tot}}(0)\notag\\
&(I\otimes\hat{\Pi}_{q_0})\;U^\dagger(t,0)\;(I\otimes\hat{\Pi}_{q_t})] \;e^{i\chi(q_t-q_0)}\Biggl], \label{cumurant_generating func}
\end{align}
where $\chi$ is often called the counting field.
The $n$th-order cumulant is given by
\begin{align}
\langle\left.\Delta q^n\rangle_t=\frac{\partial^nS_t(\chi)}{\partial(i\chi)^n}\right|_{\chi=0}.
\end{align}

\quad
In the case of the present problem, we need the energy of the environmental system as the physical quantity $Q$.
For this purpose, we set $Q=H_B$ and define the projective measurement $\{\hat{\Pi}_{e}=\ket{e}\bra{e}\}$ using the basis $\ket{e}$ of energy eigenstates.
Here, $e_0$ and $e_t$ are the measurement results at time 0 and $t$, respectively.
The projection operators are $\hat{\Pi}_{e_0}=\ket{e_0}\bra{e_0}$ and $\hat{\Pi}_{e_t}=\ket{e_t}\bra{e_t}$ using eigenstates $\ket{e_0}$ and $\ket{e_t}$, respectively.
In this particular case, since the initial state is $\rho(0)=\rho_S(0)\otimes e^{-\beta H_B}/Z$, and hence $[\rho_S(0)\otimes e^{-\beta H_B}/Z\;,\;I\otimes\hat{\Pi}_{e_0}]=0$, the following relation holds:
\begin{align}
\sum_{e_0}e^{-i\chi e_0}\hat{\Pi}_{e_0}\rho_{\mathrm{tot}}(0)\hat{\Pi}_{e_0}&=\sum_{e_0}e^{-i\chi e_0}\rho_{\mathrm{tot}}(0)\ket{e_0}\bra{e_0}\notag\\
&=e^{-i\frac{\chi}{2}H_B}\rho_{\mathrm{tot}}(0)e^{-i\frac{\chi}{2}H_B}.
\end{align}
By using this relation and the cyclic permutation of the arguments of $\mathrm{Tr}_{SB}[\cdot]$, we reduce the cumulant generation function (\ref{cumurant_generating func}) to
\begin{align}
S_t(\chi)=&\ln\Biggl[\sum_{e_0,e_t} \mathrm{Tr_{SB}}[\;(I\otimes\hat{\Pi}_{e_t})\;U(t,0)\;(I\otimes\hat{\Pi}_{e_0})\;\rho_{\mathrm{tot}}(0)\notag\\
&(I\otimes\hat{\Pi}_{e_0})\;U^\dagger(t,0)\;(I\otimes\hat{\Pi}_{e_t})] \;e^{i\chi(e_t-e_0)}\Biggl]\notag\\
=&\mathrm{Tr_{SB}}[\;e^{i\frac{\chi}{2}H_B}\;U(t,0)\;e^{-i\frac{\chi}{2}H_B}\rho_{\mathrm{tot}}(0)\notag\\
&e^{-i\frac{\chi}{2}H_B}\;U^\dagger(t,0)\;e^{i\frac{\chi}{2}H_B}]\notag\\
\equiv &\mathrm{Tr_{SB}}[\;U_{\chi/2}(t,0)\;\rho_{\mathrm{tot}}(0)\;U^\dagger_{-\chi/2}(t,0)\;],
\end{align}
where
\begin{align}
U_{\chi/2}(t,0)=e^{i\frac{\chi}{2}H_B}\;U(t,0)\;e^{-i\frac{\chi}{2}H_B},\\
U^\dagger_{-\chi/2}(t,0)=e^{-i\frac{\chi}{2}H_B}\;U^\dagger(t,0)\;e^{i\frac{\chi}{2}H_B}.
\end{align}
We then define a new state of the total system as
\begin{align}
\rho_\chi(t)=U_{\chi/2}(t,0)\;\rho_{\mathrm{tot}}(0)\;U^\dagger_{-\chi/2}(t,0).
\end{align}
By tracing out the information of the bath from this total density matrix, we derive the master equation for $\mathrm{Tr}_B\rho_\chi(t)$. 
We calculate this master equation in the counting field, and thereby obtain $S_t(\chi)=\mathrm{Tr}_{SB}\rho_\chi(t)$.
We thus obtain eq.~(\ref{NM_EB}) as the first-order cumulant of $H_B$:
\begin{align}
\Delta E_B(t)=\langle\left.\Delta H_B(t)\rangle=\frac{\partial S_t(\chi)}{\partial(i\chi)}\right|_{\chi=0}.
\end{align}

\section{Markovian dynamics}
\label{Markovian dynamics}
In this Appendix, we consider the behavior of the quantum Otto cycle when the heat exchange between the qubit and the bath are described by Markovian dynamics.
In describing this process, we can use the GKSL (Gorini-Kossakowski-Sudarshan-Lindblad) master equation under the Born-Markov approximation \cite{Gorini76, Lindblad1976}.
We can actually derive this master equation by taking the long-time limit $t \rightarrow \infty$ in the time-evolution super-operator in eq.~(\ref{master_eq}) \cite{Guarnieri16}, and hence we have \cite{Shirai21};
\begin{align}
\rho_{m,00}^{M,\mu}(t)=&\frac{1+n(\omega_\mu)}{1+2n(\omega_\mu)}+\left[\rho_{m,00}^\mu(0)-\frac{1+n(\omega_\mu)}{1+2n(\omega_\mu)}\right]\notag\\
&\mathrm{exp}\left[-2\pi J(\omega_\mu)[1+2n(\omega_\mu)]t\right]
\end{align}
with $m=0,\;1$, and $\mu=h,\;c$. 
Here, $n(\omega_\mu)$ is the Bose distribution $n(\omega_\mu)=\left(e^{\omega_\mu/T^\mu}-1\right)^{-1}$.
In order to employ the Markov approximation to describe the limit cycle, we use
\begin{align}
\rho_{11}^{M,\mu}(t^h,t^c)\equiv P^\mu\rho_{0,11}^{M,\mu}(t^\mu)+(1-P^\mu)\rho_{1,11}^{M,\mu}(t^\mu),
\end{align}
where $P^\mu$ is the same as defined in eq.~(\ref{large_p}) except that we replace the $\rho_{m,\nu\nu}^\mu(t^\mu)$ with $\rho_{m,\nu\nu}^{M,\mu}(t^\mu)$.

\bibliography{Bib_ish}
\end{document}